# Analysis of the MICCAI Brain Tumor Segmentation - Metastases (BraTS-METS) 2025 Lighthouse Challenge: Brain Metastasis Segmentation on Pre- and Post-treatment MRI

Nazanin Maleki[1,*,α,β,δ,γ], Raisa Amiruddin[1,*,α,δ,χ], Ahmed W. Moawad[188,*], Nikolay Yordanov[3,*,χ,δ,κ,], Athanasios Gkampenis[4,α,δ,κ], Pascal Fehringer[5,α,δ,κ], Fabian Umeh[6,α], Crystal Chukwurah[7, α], Fatima Memon [71,72,73,η,ρ], Bojan Petrovic[75,η,ρ], Basimah Albalooshy[1,ρ], Justin Cramer[65,η,ρ], Mark Krycia[71,η,ρ], Elizabeth B. Shrickel[80,ρ], Ichiro Ikuta[65,θ,ρ], Gerard Thompson [77,78,η,ρ], Lorenna Vidal[1,ρ], Vilma Kosovic [18,ρ], Adam E. Goldman-Yassen [70,ρ], Virginia Hill[52,ρ], Tiffany Y. So[17,ρ], Sedra Mhana[162],[χ], Albara Alotaibi[14,χ], Nathan Page[5,χ], Prisha Bhatia[15,χ], Melisa S. Guelen[189,χ], Yasaman Sharifi[19,χ], Marko Jakovljevic[1,χ], Salma A. S. Abosabie [9,16,190,χ], Sara A. Abosabie[9,133,χ], Mohanad Ghonim[162,κ,χ], Mohamed Ghonim[162,κ,χ], Amirreza Manteghinejad[1,α,χ], Ujjwal Baid [186,α,β], Anastasia Janas [9,α,β,δ], Kiril Krantchev[133,α,β,ϵ,δ], Divya Ramakrishnan[9,*,α,β], Rachit Saluja[1,11,*,α,β,γ], Leon Jekel [21,*,α,β], Maruf Adewole[22,α], Jake Albrecht[23,α], Udunna Anazodo[22, 24,α], Sanjay Aneja[25,α], Syed Muhammad Anwar[26,α], Timothy Bergquist[27,α], Veronica Chiang[28,α], Verena Chung[23,α], Gian Marco Conte[27,α], Farouk Dako[29,α], James Eddy[23,α], Ivan Ezhov[30,α], Nastaran Khalili[31,α], Keyvan Farahani[32,α], Juan Eugenio Iglesias[33,α], Zhifan Jiang[34,α], Elaine Johanson[35,α] , Anahita Fathi Kazerooni [31,36,37,α], Florian Kofler[38,α], Dominic LaBella[39,α], Koen Van Leemput[40,α], Hongwei Bran Li[33,α], Marius George Linguraru[26,41,α], Xinyang Liu[34,α], Zeke Meier[42,α], Bjoern H Menze[43,α], Harrison Moy[9,α,β,ϵ], Klara Osenberg[9,α,β], Marie Piraud[44,α], Zachary Reitman[39,α], Russell Takeshi Shinohara[45,α], Chunhao Wang[39,α], Benedikt Wiestler[38,α], Walter Wiggins[46,α], Umber Shafique[47,α,η], Klara Willms[9,β], Arman Avesta [8,9,β], Khaled Bousabarah[49,β,ϵ], Satrajit Chakrabarty[50,51,β], Nicolo Gennaro[52,β], Wolfgang Holler[49,β,ϵ], Manpreet Kaur[53,β,ϵ], Pamela LaMontagne[54,β], MingDe Lin[55,β,ϵ], Jan Lost[56,β,ϵ], Daniel S. Marcus[54,β], Ryan Maresca[25,β,ϵ], Sarah Merkaj[57,β,ϵ], Gabriel Cassinelli Pedersen[58,β,ϵ], Marc von Reppert[59,β,ϵ], Aristeidis Sotiras[54,60,β], Oleg Teytelboym[2,β], Niklas Tillmans [61,β,ϵ], Malte Westerhoff[49,β,ϵ], Ayda Youssef[62,β], Devon Godfrey[39,β], Scott Floyd[39,β], Andreas M. Rauschecker[63,β], Javier Villanueva-Meyer[63,β], Irada Pflüger[64,β], Jaeyoung Cho[64,β], Martin Bendszus[64,β], Gianluca Brugnara[64,β], Gloria J. Guzman Perez-Carillo[66,η], Derek R. Johnson[27,η], Anthony Kam[67,η], Benjamin Yin Ming Kwan[68,η], Lillian Lai[69,η], Neil U. Lall[70,η], Satya Narayana Patro[74,η], Lei Wu[79,η], Anu Bansal[81,θ], Frederik Barkhof[82,83,θ], Cristina Besada[84,θ], Sammy Chu[79,θ], Jason Druzgal[85,θ], Alexandru Dusoi[86,θ], Luciano Farage[87,θ], Fabricio Feltrin[88,θ], Amy Fong[89,θ], Steve H. Fung [90,θ], R. Ian Gray[91,θ], Michael Iv[92,θ], Alida A. Postma[93,94,θ], Amit Mahajan[9,θ], David Joyner[85,θ], Chase Krumpelman[52,θ], Laurent Letourneau-Guillon[95,θ], Christie M. Lincoln[96,θ], Mate E. Maros[97,θ], Elka Miller[98,θ]. Fanny Morón[99,θ], Esther A. Nimchinsky[100,θ], Ozkan Ozsarlak[101,θ], Uresh Patel[102,θ], Saurabh Rohatgi[48,θ], Atin Saha[103,104,θ], Anousheh Sayah[105,θ], Eric D. Schwartz[106,107,θ], Robert Shih[108,θ], Mark S. Shiroishi[109,θ], Juan E. Small[110,θ], Manoj Tanwar[111,θ], Jewels Valerie[112,θ], Brent D. Weinberg[113,θ], Matthew L. White[114,θ], Robert Young[103,θ], Vahe M. Zohrabian[115,θ], Aynur Azizova[116,θ], Melanie Maria Theresa Brüßeler[53,κ], Abdullah Okar[118,κ], Luca

Pasquini[103,κ], Gagandeep Singh[120,κ], Nico Sollmann[121,122,123,κ], Theodora Soumala[185,186,κ], Mahsa Taherzadeh[124,κ], Philipp Vollmuth[64,125,β,γ], Martha Foltyn-Dumitru[64,β,γ], Ajay Malhotra[9,β,γ], Francesco Dellepiane[126,γ], Víctor M. Pérez-García[129,γ], Hesham Elhalawani[130,γ], Maria Correia de Verdier[131,132,γ], Sanaria Al Rubaiey[133,λ], Rui Duarte Armindo[134,λ], Kholod Ashraf[62,λ], Moamen M. Asla[135,λ], Mohamed Badawy[136,λ], Jeroen Bisschop[137,λ], Nima Broomand Lomer[138,λ], Jan Bukatz[133,λ], Jim Chen[139,λ], Petra Cimflova[140,λ], Felix Corr[141,λ], Alexis Crawley[142,λ], Lisa Deptula[143,λ], Tasneem Elakhdar[62,λ], Islam H. Shawali[62,λ], Shahriar Faghani[27,λ], Alexandra Frick[144,λ], Vaibhav Gulati [145,λ], Muhammad Ammar Haider[146,λ], Fátima Hierro[147,λ], Rasmus Holmboe Dahl[148,λ], Sarah Maria Jacobs[149,λ], Kuang-chun Jim Hsieh[99,λ], Sedat G. Kandemirli[69,λ], Katharina Kersting[133,λ], Laura Kida[133,λ], Sofia Kollia[150,λ], Ioannis Koukoulithras[151,λ], Xiao Li[113,λ], Ahmed Abouelatta[62,λ], Aya Mansour[62,λ], Ruxandra-Catrinel Maria-Zamfirescu[133,λ], Marcela Marsiglia[152,λ], Yohana Sarahi Mateo-Camacho[153,λ], Mark McArthur[154,λ], Olivia McDonnell[155,λ], Maire McHugh[156,λ], Mana Moassefi[157,λ], Samah Mostafa Morsi[96,λ], Alexander Munteanu [158,λ], Khanak K. Nandolia[159,λ], Syed Raza Naqvi[160,λ], Yalda Nikanpour[161,λ], Mostafa Alnoury[162,λ], Abdullah Mohamed Aly Nouh[163,λ], Francesca Pappafava[164,λ], Markand D. Patel[165,λ], Samantha Petrucci[63,λ], Eric Rawie[166,λ], Scott Raymond[167,λ], Borna Roohani[118,λ], Sadeq Sabouhi[168,λ], Laura M. Sanchez Garcia[169,λ], Zoe Shaked[133,λ], Pokhraj P. Suthar[170,λ], Talissa Altes[171,λ], Edvin Isufi[171,λ], Yaseen Dhemesh[172,λ], Jaime Gass[171,λ], Jonathan Thacker[171,λ], Abdul Rahman Tarabishy[173,λ], Benjamin Turner[174,λ], Sebastiano Vacca[175,λ], George K. Vilanilam[174,λ], Daniel Warren[172,λ], David Weiss[176,λ], Fikadu Worede[1,λ], Sara Yousry[62,λ], Wondwossen Lerebo[1,μ], Alejandro Aristizabal[177,178,π], Alexandros Karargyris[177,π], Hasan Kassem[177,π], Sarthak Pati[8,177,179,π], Micah Sheller[177,180,π], Katherine E. Link[181,α,β], Evan Calabrese[182,α,β], Nourel Hoda Tahon[171,α,β], Ayman Nada[171,α,β], Jeffrey D. Rudie[132,184,α,β,η,ϕ], Janet Reid[1], Kassa Darge[1], Aly H. Abayazeed[92,γ], Philipp Lohmann[127,128,γ], Yuri S. Velichko[52,α,β], Spyridon Bakas[8,47,183,α,β,ϕ], Mariam Aboian[1,α,β,η,ϕ,γ,†]

1. Department of Radiology, Children's Hospital of Philadelphia, Philadelphia, PA, USA
2. Trinity Health Mid Atlantic Hospitals, Darby, PA, USA
3. Department of Neurointensive Care, Multiprofile Hospital for Active Treatment in Neurology and Psychiatry "St. Naum", Sofia, Bulgaria
4. Department of Neurosurgery and Neurotechnology, University of Tübingen, Tübingen, Germany
5. Faculty of Medicine, Jena University Hospital, Friedrich Schiller University Jena, Jena, Germany
6. Teesside University, Middlesbrough, United Kingdom
7. Yale School of Medicine, New Haven, Connecticut, USA
8. Division of Computational Pathology, Department of Pathology and Laboratory Medicine, Indiana University School of Medicine, Indianapolis, IN, USA
9. Department of Radiology and Biomedical Imaging, Yale School of Medicine, New Haven, CT, USA

10. Department of Electrical and Computer Engineering, Cornell University and Cornell Tech, New York, NY, USA
11. Department of Radiology, Weill Cornell Medicine, New York, NY, USA
12. Department of Radiology, Children's Hospital of Philadelphia, Philadelphia, PA, USA
13. Pritzker School of Medicine - University of Chicago, Chicago, IL, USA
14. Jordan University of Science and Technology, Irbid, Jordan
15. Mohammed Bin Rashid University of Medicine and Health Sciences, Dubai, United Arab Emirates
16. Institute of Experimental Biomedicine, University Hospital Würzburg, Würzburg, Germany
17. Department of Imaging & Interventional Radiology, Chinese University of Hong Kong, Hong Kong
18. Department of Radiology, General Hospital of Dubrovnik, Dubrovnik, Croatia
19. Department of Radiology, Iran University School of Medicine, Tehran, Iran
20. College of Medicine, Alfaisal University, Riyadh, Saudi Arabia
21. DKFZ Division of Translational Neurooncology at the WTZ, German Cancer Consortium, DKTK Partner Site,University Hospital Essen, Essen, Germany
22. Medical Artificial Intelligence Lab, Crestview Radiology, Lagos, Nigeria
23. Sage Bionetworks, Seattle, WA, USA
24. Montreal Neurological Institute, McGill University, Montreal, Canada
25. Department of Therapeutic Radiology, Yale School of Medicine, New Haven, CT, USA
26. Sheikh Zayed Institute for Pediatric Surgical Innovation, Children's National Hospital, Washington, D.C., USA
27. Department of Radiology, Mayo Clinic, Rochester, MN, USA
28. Department of Neurosurgery, Yale School of Medicine, New Haven, CT, USA
29. Center for Global Health, Perelman School of Medicine, University of Pennsylvania, PA, USA
30. Department of Informatics, Technical University Munich, Germany
31. Center for Data-Driven Discovery in Biomedicine, Children's Hospital of Philadelphia, Philadelphia, PA, USA
32. Cancer Imaging Program, National Cancer Institute, National Institutes of Health, Bethesda, MD, USA
33. Athinoula A. Martinos Center for Biomedical Imaging, Massachusetts General Hospital, Boston, MA, USA
34. Children's National Hospital, Washington, D.C., USA
35. PrecisionFDA, U.S. Food and Drug Administration, Silver Spring, MD, USA
36. Department of Neurosurgery, University of Pennsylvania, Philadelphia, PA, USA
37. Division of Neurosurgery, Children's Hospital of Philadelphia, Philadelphia, PA, USA
38. Department of Neuroradiology, Technical University of Munich, Munich, Germany
39. Department of Radiation Oncology, Duke University Medical Center, Durham, NC, USA

40. Department of Applied Mathematics and Computer Science, Technical University of Denmark, Denmark
41. Departments of Radiology and Pediatrics, George Washington University School of Medicine and Health Sciences, Washington, D.C., USA
42. Booz Allen Hamilton, McLean, VA, USA
43. Biomedical Image Analysis & Machine Learning, Department of Quantitative Biomedicine, University of Zurich, Switzerland
44. Helmholtz AI, Helmholtz Munich, Germany
45. Center for Clinical Epidemiology and Biostatistics, University of Pennsylvania, Philadelphia, PA, USA
46. Duke University School of Medicine, Durham, NC, USA
47. Department of Radiology and Imaging Sciences, Indiana University, Indianapolis, IN, USA
48. Department of Radiology, Neuroradiology, Massachusetts General Hospital, Boston, MA, USA
49. Visage Imaging, GmbH, Berlin, Germany
50. Department of Electrical and Systems Engineering, Washington University in St. Louis, St. Louis, MO, USA
51. GE HealthCare, San Ramon, CA, USA
52. Department of Radiology, Northwestern University, Feinberg School of Medicine, Chicago, IL, USA
53. Ludwig Maximilian University, Munich, Germany
54. Mallinckrodt Institute of Radiology, Washington University School of Medicine, St. Louis, MO, USA
55. Visage Imaging, Inc, San Diego, CA, USA
56. Department of Neurosurgery, Heinrich-Heine University, Moorenstrasse 5, Dusseldorf, Germany
57. University of Ulm, Ulm, Germany
58. University of Göttingen, Göttingen, Germany
59. University of Leipzig, Leipzig, Germany
60. Institute for Informatics, Data Science & Biostatistics, Washington University School of Medicine, St. Louis, MO, USA
61. Department of Diagnostic and Interventional Radiology, Medical Faculty, University Dusseldorf, Dusseldorf, Germany
62. Cairo University, Cairo, Egypt
63. Department of Radiology and Biomedical Imaging, University of California San Francisco, CA, USA
64. Department of Neuroradiology, Heidelberg University Hospital, Heidelberg, Germany
65. Department of Radiology, Mayo Clinic, Phoenix, AZ, USA

66. Neuroradiology Section, Mallinckrodt Institute of Radiology, Washington University in St. Louis, St. Louis, MO, USA
67. Loyola University Medical Center, Hines, IL, USA
68. Department of Radiology, Queen's University, Kingston, ON, Canada
69. Department of Radiology, University of Iowa Hospitals and Clinics, Iowa City, IA, USA
70. Children's Healthcare of Atlanta, Emory University School of Medicine, GA, USA
71. Carolina Radiology Associates, Myrtle Beach, SC, USA
72. McLeod Regional Medical Center, Florence, SC, USA
73. Medical University of South Carolina, Charleston, SC, USA
74. University of Arkansas Medical Center, Little Rock, AR, USA
75. NorthShore Endeavor Health, Evanston, IL, USA
76. Department of Imaging and Interventional Radiology, The Chinese University of Hong Kong, Hong Kong SAR
77. Centre for Clinical Brain Sciences, University of Edinburgh, Edinburgh, United Kingdom
78. Department of Clinical Neurosciences, NHS Lothian, Edinburgh, United Kingdom
79. Department of Radiology, University of Washington, Seattle, WA, USA
80. Department of Radiology, Ohio State University College of Medicine, Columbus, OH, USA
81. Albert Einstein Medical Center, Hartford, CT, USA
82. Amsterdam UMC, location Vrije Universiteir, Netherlands
83. University College London, United Kingdom
84. Hospital Italiano de Buenos Aires, Buenos Aires, Argentina
85. Department of Radiology and Medical Imaging, University of Virginia, Charlottesville, Virginia, USA
86. Klinikum Hochrhein, Waldshut-Tiengen, Germany
87. Centro Universitario Euro-Americana (UNIEURO), Brasília, DF, Brazil
88. Department of Radiology, University of Texas Southwestern Medical Center, Dallas, TX, USA
89. Southern District Health Board, Dunedin, New Zealand
90. Department of Radiology, Houston Methodist, Houston, TX, USA
91. University of Tennessee Medical Center, Knoxville, TN, USA
92. Department of Radiology, Stanford University, Stanford, CA, USA
93. Department of Radiology and Nuclear Medicine, Maastricht University Medical Center, Maastricht, the Netherlands
94. Mental Health and Neuroscience Research Institute, Maastricht University, Maastricht, the Netherlands
95. Centre Hospitalier de l'Universite de Montreal and Centre de Recherche du CHUM Montreal, Canada
96. Department of Neuroradiology, MD Anderson Cancer Center, Houston, TX, USA

97. Departments of Neuroradiology & Biomedical Informatics, Medical Faculty Mannheim, Heidelberg University, Mannheim, Germany
98. Department of Diagnostic and Interventional Radiology, SickKids Hospital, University of Toronto, Canada
99. Department of Radiology, Baylor College of Medicine, Houston, TX, USA
100. Department of Radiology, New Jersey Medical School, Newark, NJ, USA
101. Department of Radiology, AZ Monica, Antwerp Area, Belgium
102. Medicolegal Imaging Experts LLC, Mercer Island, WA, USA
103. Department of Radiology, Memorial Sloan Kettering Cancer Center, New York, NY, USA
104. Weill Cornell Medical College, New York, NY, USA
105. MedStar Georgetown University Hospital, Washington, D.C., USA
106. Department of Radiology, St. Elizabeth's Medical Center, Boston, MA, USA
107. Department of Radiology, Tufts University School of Medicine, Boston, MA, USA
108. Walter Reed National Military Medical Center, Bethesda, MD, USA
109. Keck School of Medicine, Los Angeles, CA, USA
110. Department of Radiology, Lahey Hospital and Medical Center, Burlington, MA, USA
111. Department of Radiology, University of Alabama, Birmingham, AL, USA
112. Department of Radiology, University of North Carolina School of Medicine, Chapel Hill, NC, USA
113. Department of Radiology and Imaging Sciences, Emory University, Atlanta, GA, USA
114. University of Nebraska Medical Center, Omaha, NE, USA
115. Northwell Health, Zucker Hofstra School of Medicine at Northwell, North Shore University Hospital, Hempstead, New York, NY, USA
116. Cancer Center Amsterdam, Imaging and Biomarkers, Amsterdam, The Netherlands
117. Department of Radiology, Ain Shams University, Cairo, Egypt
118. University of Hamburg, University Medical Center Hamburg-Eppendorf, Hamburg, Germany
119. Department of Radiology, Iran University of Medical Sciences, Tehran, Iran
120. Columbia University Irving Medical Center, New York, NY, USA
121. Department of Diagnostic and Interventional Radiology, University Hospital Ulm, Ulm, Germany
122. Department of Diagnostic and Interventional Neuroradiology, School of Medicine, Klinikum rechts der Isar, Technical University of Munich, Munich, Germany
123. TUM-Neuroimaging Center, Klinikum rechts der Isar, Technical University of Munich, Munich, Germany
124. Department of Radiology, Arad Hospital, Tehran, Iran
125. Department of Medical Image Computing, German Cancer Research Center (DKFZ), Heidelberg, Germany

126. Functional and Interventional Neuroradiology Unit, Bambino Gesù Children's Hospital, Rome, Italy
127. Institute of Neuroscience and Medicine (INM-4), Research Center Juelich, Juelich, Germany
128. Department of Nuclear Medicine, University Hospital RWTH Aachen, Aachen, Germany
129. Mathematical Oncology Laboratory & Department of Mathematics, University of Castilla-La Mancha, Spain
130. Department of Radiation Oncology, Brigham and Women's Hospital, Harvard Medical School, Boston, MA, USA
131. Department of Surgical Sciences, Section of Neuroradiology, Uppsala University, Sweden
132. Department of Radiology, University of California San Diego, CA, USA
133. Charité-Universitätsmedizin Berlin (Corporate Member of Freie Universität Berlin, Humboldt-Universität zu Berlin,und Berlin Institute of Health), Berlin, Germany
134. Department of Neuroradiology, Western Lisbon Hospital Centre (CHLO), Portugal
135. Zagazig University, Zagazig, Egypt
136. Diagnostic Radiology Department, Wayne State University, Detroit, MI
137. Institute of Diagnostic and Interventional Radiology, University Hospital Zurich, University of Zurich, Zurich, Switzerland.
138. Faculty of Medicine, Guilan University of Medical Sciences, Rasht, Iran
139. Department of Radiology/Division of Neuroradiology, San Diego Veterans Administration Medical Center/UC San Diego Health System, San Diego, CA, USA
140. Department of Radiology, University of Calgary, Calgary, Canada
141. EDU Institute of Higher Education, Villa Bighi, Chaplain's House, Kalkara, Malta
142. Bay Imaging Consultants, Walnut Creek, CA, USA
143. Ross University School of Medicine, Bridgetown, Barbados
144. Department of Neurosurgery, Vivantes Klinikum Neukölln, Berlin, Germany
145. Mercy Catholic Medical Center, Darby, PA, USA
146. C.M.H. Lahore Medical College, Lahore, Pakistan
147. Neuroradiology Department, Pedro Hispano Hospital, Matosinhos, Portugal
148. Department of Radiology, Copenhagen University Hospital - Rigshospitalet, Copenhagen, Denmark
149. Rijnstate Hospital, Arnhem, Netherlands
150. National and Kapodistrian University of Athens, School of Medicine, Athens, Greece
151. Department of Neurosurgery, University Hospital of Ioannina, Ioannina, Greece
152. Department of Radiology, Brigham and Women's Hospital, Massachusetts General Hospital, Boston, MA, USA
153. Department of Neuroradiology, Universidad Autónoma de Nuevo Léon, Mexico

154. Department of Radiological Sciences, University of California Los Angeles, Los Angeles, CA, USA
155. Gold Coast University Hospital, Queensland Health, Australia
156. Department of Radiology Manchester NHS Foundation Trust, North West School of Radiology, Manchester, United Kingdom
157. Artificial Intelligence Lab, Department of Radiology, Mayo Clinic, Rochester, MN, USA
158. Corewell Health West, MI, USA
159. Department of Radiodiagnosis, All India Institute of Medical Sciences Rishikesh, India
160. Windsor Regional Hospital, Western University, Ontario, Canada
161. Artificial Intelligence & Informatics, Mayo Clinic, Rochester, MN, USA
162. Department of Radiology, University of Pennsylvania, PA, USA
163. Department of Radiology, Life Care Hospital, Freetown, Sierra Leone
164. Department of Medicine and Surgery, Università degli Studi di Perugia, Italy
165. Department of Neuroradiology, Imperial College Healthcare NHS Trust, London, United Kingdom
166. Department of Radiology, Michigan Medicine, Ann Arbor, MI, USA
167. Department of Radiology, University of Vermont Medical Center, Burlington, VT, USA
168. Isfahan University of Medical Sciences, Isfahan, Iran
169. Department of Radiology, The American British Cowdray Medical Center, Mexico City, Mexico
170. Rush University Medical Center, Chicago, IL, USA
171. Radiology Department, University of Missouri, Columbia, MO, USA
172. Washington University School of Medicine in St. Louis, St. Louis, MO, USA
173. Department of Neuroradiology, Rockefeller Neuroscience Institute, West Virginia University. Morgantown, WV, USA
174. Leeds Teaching Hospitals NHS Trust, Leeds, United Kingdom
175. University of Cagliari, School of Medicine and Surgery, Cagliari, Italy
176. Department of Diagnostic and Interventional Radiology and Neuroradiology, University Hospital Essen, Essen, Germany
177. MLCommons, San Francisco, CA, USA
178. Factored, Palo Alto, CA, USA
179. Center For Federated Learning in Medicine, Indiana University, Indianapolis, IN, USA
180. Intel Corporation, Hillsboro, OR, USA
181. New York University School of Medicine, New York, NY, USA
182. Department of Radiology, Duke University Medical Center, Durham, NC, USA
183. Department of Neurological Surgery, School of Medicine, Indiana University, Indianapolis, IN, USA
184. Department of Radiology, Scripps Clinic Medical Group, CA, USA
185. Center for Neurology, Academic Specialist Center, Stockholm, Sweden

186. Department of Clinical Neuroscience, Karolinska Institute, Stockholm, Sweden
187. The Wallace H. Coulter Department of Biomedical Engineering at Georgia Tech and Emory University, Atlanta, GA, USA
188. Department of Radiology, Division of Neuroradiology, Perelman School of Medicine at the University of Pennsylvania
189. Athinoula A. Martinos Center for Biomedical Imaging, Massachusetts General Hospital, Boston, MA, USA
190. Rudolf Virchow Center for Integrative and Translational Bioimaging, Julius-Maximilians-Universität Würzburg, Würzburg, Germany

* Equal First Authors
α Organizer
β Data contributors
δ International lead
ϵ Annotator
η Super Approver
θ Rest of Approvers
κ Super Annotator
λ Rest of Annotators
μ Stats
π MLCommons
ϕ Equal Senior Authors
ρ Reference dataset faculty annotator
χ Reference dataset faculty observer
† Corresponding Author — aboianm@chop.edu

AUTHOR LIST IS NOT FINAL YET AND WILL BE UPDATED AT FINAL SUBMISSION

**Abstract:** Despite continuous advancements in cancer treatment, brain metastatic disease remains a significant complication of primary cancer and is associated with an unfavorable prognosis. One approach for improving diagnosis, management, and outcomes is to implement algorithms based on artificial intelligence for the automated segmentation of both pre- and post-treatment MRI brain images. Such algorithms rely on volumetric criteria for lesion identification and treatment response assessment, which are still not available in clinical practice. Therefore, it is critical to establish tools for rapid volumetric segmentations methods that can be translated to clinical practice and that are trained on high quality annotated data. The BraTS-METS 2025 Lighthouse Challenge aims to address this critical need by establishing inter-rater and intra-rater variability in dataset annotation by generating high quality annotated datasets from four individual instances of segmentation by neuroradiologists while being recorded on video (two instances doing 'from scratch' and two instances after AI pre-segmentation). This high-quality annotated dataset will be used for testing phase in 2025 Lighthouse challenge and will be publicly released at the completion of the challenge. The 2025 Lighthouse challenge will also release the 2023 and 2024 segmented datasets that were annotated using an established pipeline of pre-segmentation, student annotation, two neuroradiologists checking, and one neuroradiologist finalizing the process. It builds upon its previous edition by including post-treatment cases in the dataset. Using these high-quality annotated datasets, the 2025 Lighthouse challenge plans to test benchmark algorithms for automated segmentation of pre-and post-treatment brain metastases (BM), trained on diverse and multi-institutional datasets of MRI images obtained from patients with brain metastases.

The data includes T1-weighed, post-contrast T1-weighed, and T2 FLAIR sequences, with T2-weighed becoming non-mandatory for the current edition of the challenge. A total of 1475 cases were released on the Synapse webpage of the challenge, with 1296 designated for training and 179 for validation. The participants will be evaluated using several metrics: Dice Score Coefficient, Normalized Surface Distance, sensitivity, specificity, and precision of the model. In conclusion, The BraTS-METS 2025 Lighthouse Challenge creates a competitive environment for the development of robust automatic segmentation algorithms for pre- and post-treatment BM. The aim is to improve lesion segmentation and facilitate a more efficient assessment of disease progression using volumetric criteria, ultimately reducing the time required for evaluation.

## I. Introduction:

Brain metastases (BM) are the most commonly diagnosed central nervous system (CNS) neoplasm in adults. Advances in cancer treatments have resulted in increased survival rates and life expectancy in cancer patients[1]. Additionally, improvements of medical imaging techniques have enabled more sensitive detection of asymptomatic lesions[2]. As a result of these factors, the incidence of brain metastases has increased worldwide. It is estimated that brain metastatic disease develops in 20 to 40 percent of all the patients with primary cancer, with these percentages varying between primary cancer types and across studies[3,2,4]. Despite advancements in treatment options, the median survival for patients with stable extracranial disease following first-line treatment of brain metastases is only 16.9 months[5]. These statistics demonstrate the significant social burden posed by brain metastatic disease and highlight the necessity for exploiting new approaches for improving the care for these patients. Given the promising capabilities of artificial intelligence and the crucial role of neuroimaging in brain tumor diagnosis and treatment response assessment, intersecting those fields could be the cornerstone for improving the outcomes for patients with brain tumors[6,7].

The treatment of brain metastatic disease encompasses different methods, including surgical resection, radiotherapy, whole-brain radiation, systemic targeted therapy or immunotherapy[8]. Stereotactic radiosurgery (SRS) remains the mainstay of treatment for brain metastases, particularly effective for patients with a relatively small number of metastases (one to four), and for lesions that do not cause prevalent mass effect or located at areas that do not allow surgical intervention [8,9,10]. Recent advances of SRS have expanded the indications for applying this method, allowing it to be used for patients with multiple metastases, or for lesions that are close to eloquent anatomical regions or radiotherapy-sensitive tissues[9,10]. Moreover, improvements in systemic therapy have led to an increase in life expectancy for these patients, enabling multiple courses of SRS to be administered[11]. The presence of numerous metastases treated at various time points, along with the potential occurrence of new lesions, makes it extremely difficult for radiologists to evaluate the treatment response for each lesion and to manually quantify the change in volume. This quantification is of great importance since an increase in volume may necessitate closer monitoring, additional imaging, and possible alterations in treatment course[11]. The lack of standardization for the imaging modality used and frequency of assessments, along with the questionable criteria for determining treatment response, further complicates the evaluation of brain metastases cases. Additionally, clinicians are faced with the tedious task of providing a personalized treatment approach that aims to minimize the toxic effects of treatment and maximize the chances of cure[7].

There is a significant potential for Artificial intelligence (AI) tools, such as machine learning (ML) or deep learning (DL), to address the challenges mentioned above. These tools can deliver accurate volumetric assessment of brain metastases, aiding in the formulation of treatment strategy. They also have the capacity to play a vital role in monitoring the size of lesions that have undergone SRS treatment[7,11].

The BraTS-METS challenge conducted in 2023 marked an important first step toward establishing AI as a valuable tool that could improve radiology practice, patient clinical decision-making, and patient outcomes. Robust algorithms were created for automatic segmentation of brain metastases on various MRI sequences, including T1-weighed, post-contrast T1-weighed, T2-weighed and FLAIR[12]. One of the critical lessons learned from 2023 and 2024 challenges, including from the literature[13,14,15], and discussion with researchers at MICCAI 2022 and 2023, was that there is significant inter- and intra-rater variability in annotation of datasets. The BraTS-METS 2025 Lighthouse Challenge aims to address the critical need for reliable ground truth metastases annotations by establishing inter-rater and intra-rater variability in dataset annotation by generating high quality annotated datasets from four individual instances of segmentation by neuroradiologists while being recorded on video (two instances doing ‘from scratch’ and two instances after AI pre-segmentation). These high-quality annotated datasets will be used for testing in 2025 Lighthouse Challenge and will provide for the first time new understanding of expectations that reference standard requires. In addition, the BraTS-METS 2025 Lighthouse Challenge will build upon the strong foundation established in previous years and will provide high quality annotated data of pre- and post-treatment brain metastases for development of novel segmentation algorithms by the competitors. All the steps mentioned above align with the long-term goal of establishing an open-source brain metastasis consortium for research collaboration, aimed at creating a repository of imaging studies of pre- and post-treatment metastasis with accompanying data regarding tumor type, treatment modality, and patient outcomes.

## II. Materials and methods:

### 1. Data description

The most sensitive method for detecting and evaluating brain metastases is Magnetic Resonance Imaging (MRI), making it the designated imaging modality for this challenge. The 2025 BraTS-METS dataset contains multiparametric MRI scans obtained retrospectively from various institutions across the US and internationally. This strategy ensures a diverse dataset in terms of image quality and acquisition parameters, reflecting the global variability in imaging practices. Only MRI studies of patients diagnosed with brain metastases were included. In contrast to the dataset released for the 2023 and 2024 editions of the BraTS-METS challenge which included only images obtained in the pre-treatment period, this year's dataset also contains images acquired in the post-treatment period.

The contrast-enhanced T1-weighed imaging sequence provides the highest sensitivity for detecting brain metastases and is regarded as the gold standard for BM segmentation[4,16]. FLAIR is the most reliable sequence for edema segmentation. Therefore, to be included in the challenge, a case had to comprise the following sequences: pre-contrast T1-weighed imaging, post-contrast T1-weighed imaging, and FLAIR. For the 2025 edition of the challenge, T2-weighed imaging is no longer mandatory, as was the case for the 2024 challenge. Cases with significant imaging artifacts were excluded from the dataset.

Aligning with the conventional data split strategy in machine learning, 70% of the cases were allocated to the training set, 10% to the validation set, and 20% to the testing set. In addition to the standard testing set, the multi-annotated dataset of 75 subjects will be added to the testing set. The training set is provided with accompanying ground truth segmentations. The validation set is provided without any segmentation masks so that the participants can test check their model in an unbiased manner. The testing set is hidden from the participants, but the multi-annotated 75 subject dataset will be released to the public after the completion of the challenge. This dataset comprises 75 brain metastases cases that have undergone 4 stages of annotation by two independent board-certified neuroradiologists, with an intermission of 7 days between each of those stages for suppression of image recall. In summary, every case from this additional dataset will have 8 corresponding annotations, four created manually from scratch, and the remaining four generated by refining nnU-Net-based automatic segmentations of the cases.

### 2. Pre-processing protocol

All the mpMRI cases within the BraTS-METS dataset have undergone standardized pre-processing, based on Cancer Imaging Phenomics Toolkit (CaPTk) and Federated Tumor Segmentation (FeTS) Tool[17,18]. The first step of the pre-processing protocol involved converting the Digital Imaging and Communication in Medicine (DICOM) files to Neuroimaging Informatics Technology Initiative (NIfTI) file format. During this conversion, all protected health information and metadata supplementing the DICOM image are removed.

To ensure proper alignment of the separate masks (enhancing tumor, non-enhancing tumor core, edema, etc.) obtained during the segmentation process, it was necessary for the images corresponding to different acquisition sequences to be co-registered in the same spatial framework[16]. Factors such as tumor progression, edema, resection cavities and midline shift can change the coordinates of certain metastatic lesions within the brain. Consequently, longitudinal studies must also be co-registered so that the model can accurately identify and track the same lesion across images acquired at different time points along the treatment course[19]. Therefore, as the second step of the pre-processing pipeline, the cases have been registered to the same anatomical atlas (except UCSD and UCSF datasets), specifically the SRI24 multichannel atlas of normal adult brain. However, image registration may lead to some small metastases being missed due to interpolation. In addition, it is more intuitive for the interpreting radiologist to evaluate studies without any interpolation. For this reason, all the cases that haven't been registered in SRI24 space, are provided in their native space. It is the goal for future challenges to transition all images to native space.

According to the National Comprehensive Cancer Network guidelines, it is recommended to obtain T1-weighed contrast-enhanced MRI with 1-mm thick slices within 14 days of starting treatment with stereotactic radiosurgery[20]. For the dataset to comply with the current clinical recommendations, all the images were resampled to a uniform isotropic resolution ($1mm^3$) as the third step of pre-processing.

The last step of the pre-processing pipeline involved skull-stripping and anonymization of the available images. The utilized approach was developed by Thakur et al. and is based on a deep learning (DL) model trained on multi-modal data. This allows the model to be effectively applied to any MRI sequence and to reliably perform on multi-institutional data[21].

3. Tumor annotation protocol

For the BraTS 2025 Lighthouse Challenge, a 4-label system for image annotation is used. The labels are the following:

- *Label 1, Non-enhancing tumor core (NETC):* This label represents all portions of the tumor core that do not enhance and are surrounded by enhancing tumor. This label corresponds to the necrotic core of the tumor.
- *Label 2, Surrounding non-enhancing FLAIR hyperintensity (SNFH):* This label corresponds to the hyperintense signal on T2-FLAIR sequences. It identifies tissue that is infiltrated by the infiltrative non-enhancing tumor, or that is edematous because of vasogenic edema. Prior infarcts or microvascular ischemic white matter changes that exhibit high T2-FLAIR signal are not labelled as SNFH.

- *Label 3, Enhancing Tumor (ET):* This label marks the portions of the tumor that take up contrast on postcontrast T1-weighed images. Adjacent blood vessels, bleeding or intrinsic T1 hyperintensity are not included in this class.
- *Label 4, Resection Cavity (RC):* Delineates the resection region within the brain in post-treatment cases. Majority of the available cases do not have this segmentation.

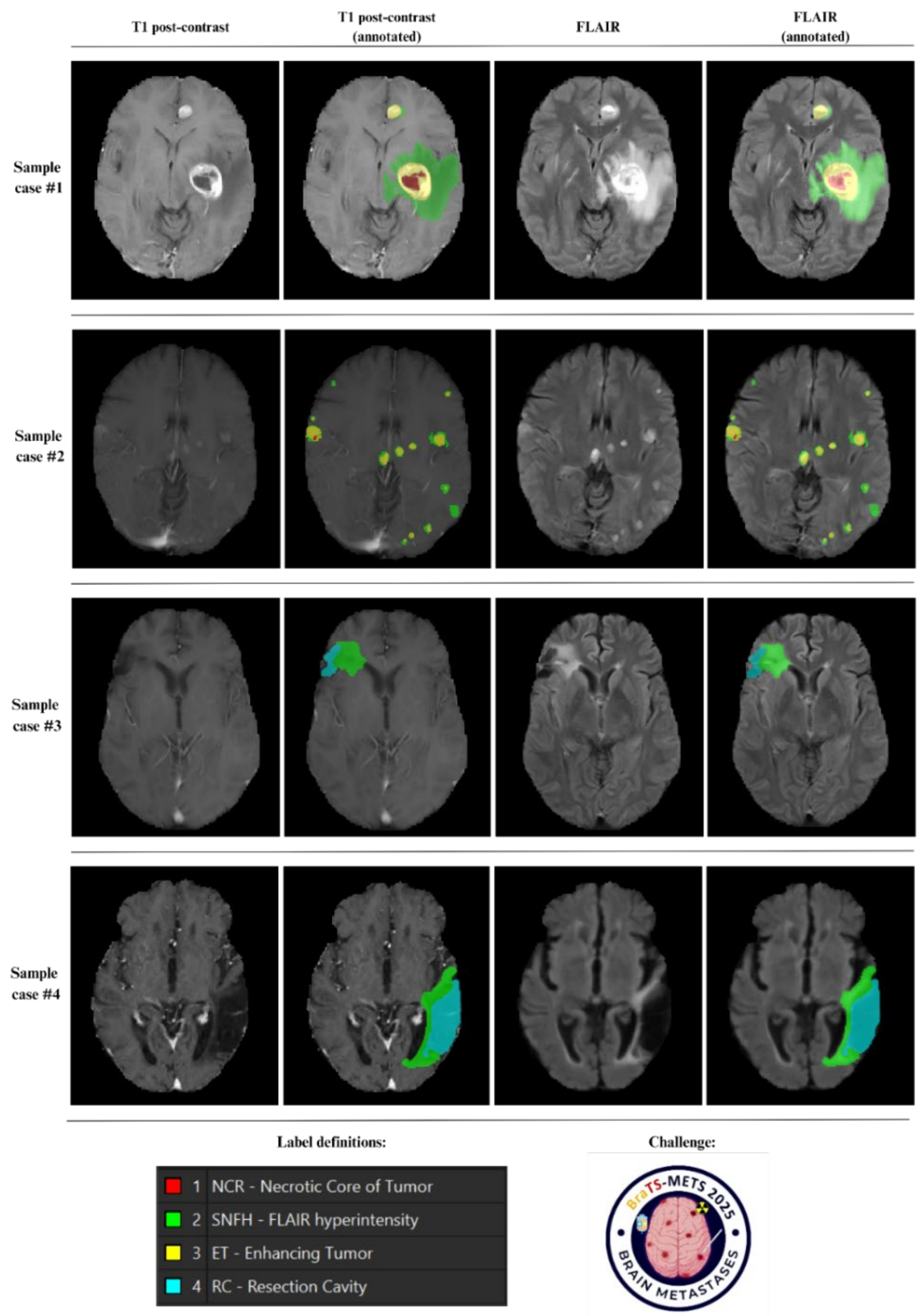

*Figure 1.* Examples of annotated pre-treatment (Sample cases 1 and 2) and post-treatment (Sample cases 3 and 4) MRI images of patients with brain metastatic disease.

To assess and evaluate the brain tumor lesions on the imaging data in a reliable and reproducible way, a standardized protocol for defining the tumor subregions was necessary. For the 2025 edition of the challenge, the following sub-regions were defined: enhancing tumor, tumor core and whole tumor. More details are depicted in Table 1. Some examples are provided in Figure 1.

| **Component** | **Label** |
|---|---|
| Enhancing Tumor (ET) | Label 2 (ET) |
| Tumor Core (TC) | Label 2 (ET) + Label 3 (SNFH) |
| Whole Tumor (WT) | Label 2 (ET) + Label 3 (SNFH) + Label 1 (NETC) |

*Table 1.* Regions of interest (ROI) for the BraTS-METS 2025 Lighthouse Challenge

The complete extent of the disease is defined by the volume of both the tumor core and the surrounding edema and/or infiltration. In some instances, the perilesional edema from two or more separate lesions may overlap. Therefore, enhancing tumor (ET) and whole tumor (WT) are treated as distinct entities.

The cases within the dataset have been annotated to establish the reference standard labels. A five-step annotation pipeline was employed for this purpose (Figure 2). The first step involved automatic pre-segmentation of the available images using approaches based on nnU-Net. The University of California, San Francisco Brain Metastases Stereotactic Radiosurgery MRI Dataset (UCSF-BMSR) was used to train the nnU-Net tool to generate GT segmentations of the enhancing tumor (ET). These segmentations were then fused with the labels for NETC and SNFH, which were generated by nnU-Net based on the BraTS 2021 pre-treatment glioma dataset[22,23]. Additionally, the AURORA multicenter study and the Heidelberg University Hospital datasets were applied separately for training the nnU-Net tool to generate the labels for SNFH and TC (ET+NETC)[24,25]. The labels generated for ET were fused using a minority voting algorithm to ensure that no small metastases are excluded. The labels for SNFH were fused using the STAPLE fusion algorithm. The label for NETC was created using only one of the three pre-segmentation approaches mentioned above, therefore it was overlaid on the ET and SNFH labels.

In the second step of the annotation protocol, all the acquired pre-segmentations were provided to a group of volunteer medical student annotators, and board-certified neuroradiologists. Their task was to correct any mistakes and refine the pre-segmentations. All the annotations performed by the students were reviewed by a cohort of 52 neuroradiology faculty. Cases that lacked completeness were returned to the students for further refinement. The annotations accepted by the neuroradiologist underwent Quality Control, which is the third step of the annotation pipeline. During this step, any voxels outside the brain parenchyma, as well as random voxels not associated with any lesion, were removed from the segmentation mask. It was ensured that all the sequences were located in their corresponding folders within the dataset and that all the images were in the same space and orientation. The fourth step of the annotation pipeline involved a second neuroradiologist who conducted a secondary review to ensure the segmentations met the required quality standards. The fifth step of the pipeline was executed by a senior neuroradiologist who carefully reviewed all the segmentations to ensure that high standard annotations are consistent throughout the dataset and finalized the annotation process. The UCSD dataset is new for the 2025 challenge and was annotated by one board-certified neuroradiologist, separate from the above described five step pipeline. In contrast to the other datasets, the UCSD dataset consists of longitudinal, progressive data that represents the realistic follow-up scenario in which a patient is monitored with serial imaging studies.

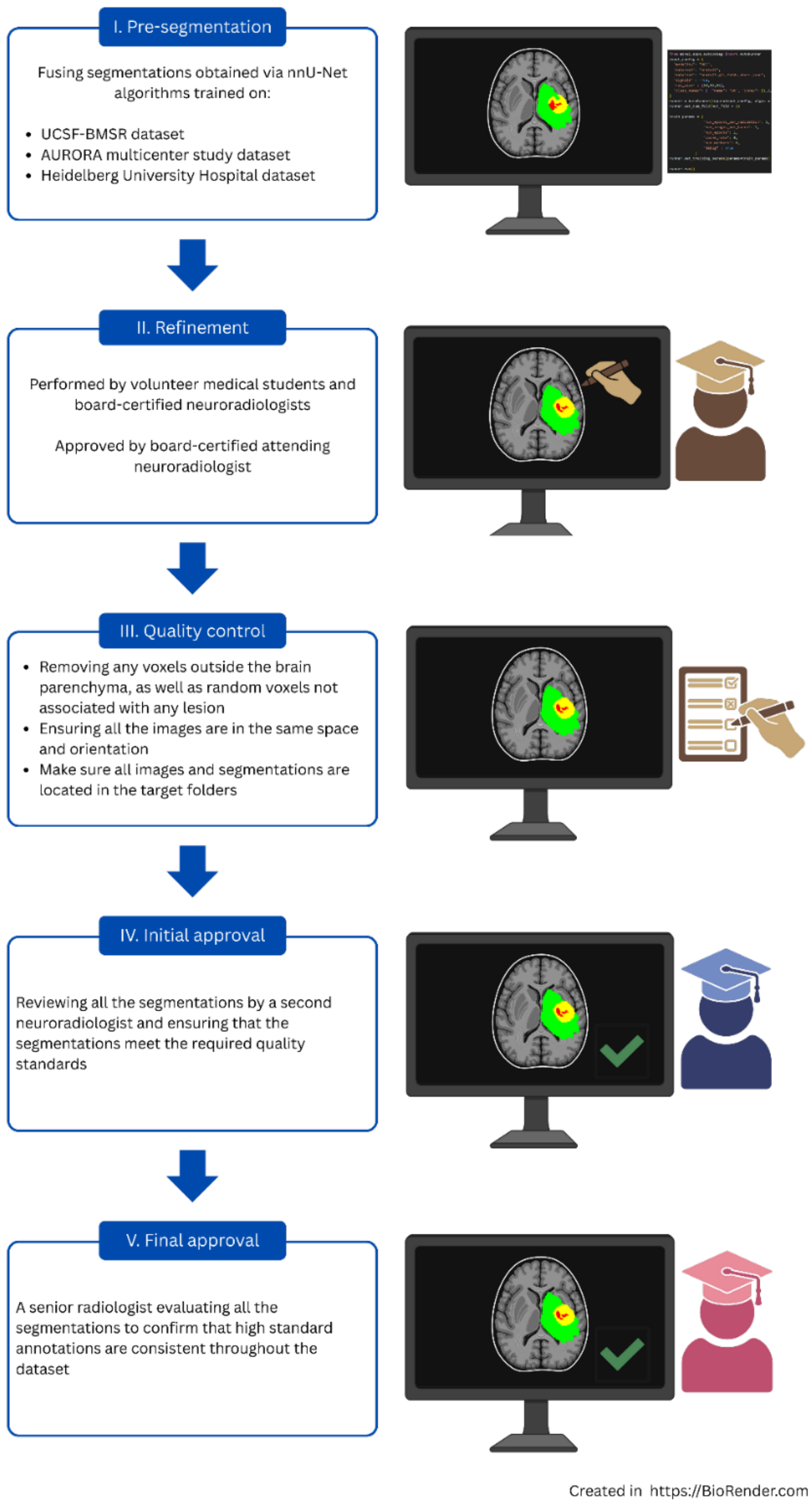

*Figure 2.* Graphical representation of the five-step annotation pipeline utilized for the challenge.

4. Performance evaluation

To create their brain tumor segmentation model, challenge participants are asked to use MedPerf, an open-source framework provided by MLCommons, an AI engineering consortium. This proposed framework, namely the Generally Nuanced Deep Learning Framework (GaNDLF), allows participants with limited computational experience to train deep learning (DL) models on radiology data. Additionally, it allows them to test their methods and evaluate their performance on a wide array of medical datasets[26]. Participants should download a Docker container, namely the MLCube, from the challenge webpage, which is supported by the Synapse platform. Before submitting their model to the BraTS webpage, they should package it within this Docker container.

Selected metrics used to validate model performance include Dice Similarity Coefficient (DSC) applied to measure the spatial overlap between the model's predictions and the reference standard derived from manual annotation of the data[27], Normalized Surface Distance (NSD) used to measure the overlap between the lesion boundaries set by the model and those corresponding to the ground truth. Sensitivity, specificity and precision of the model were also applied. The overall ranking of each participating team will be based on summation of their ranks for all the metrics mentioned above. Statistical significance testing of the ranking strategy will also be employed. The metrics and ranking strategy used the recommendations based on the DELPHI method for image analysis research proposed by Reinke et al.[28,29]

## III. Results:

A total of 1778 multi-parametric pre- and post-treatment cases were included in the BraTS-METS 2025 Lighthouse Challenge dataset. These cases were provided by multiple institutions: Duke University, Missouri University, Wahington University in St. Louis, Yale University, University of California, San Francisco (UCSF), University of California, San Diego (UCSD), Northwestern University, and the National Cancer Institute in Egypt (Table 2 & Figure 3).

Within the BraTS-METS 2025 dataset, there are a total of 732 post-treatment brain metastases cases and 1046 pre-treatment cases. The UCSF and UCSD datasets together account for 1,272 cases, which were registered in native space. The remaining 506 cases were registered using the SRI24 anatomical atlas. The training portion of the BraTS-METS 2025 dataset comprises 1,296 cases, while the validation and testing portions include 179 and 303 cases, respectively.

| Dataset: | Training | Validation | Testing | Treatment stage: | Post-treatment cases, n = | Registered in: |
|---|---|---|---|---|---|---|
| Duke | 37 | 15 | 30 | pre-treatment | 0 | SRI24 space |
| NCI | 35 | 0 | 1 | pre-treatment | 0 | SRI24 space |
| Missouri | 22 | 25 | 35 | pre-treatment | 0 | SRI24 space |
| WashU | 39 | 2 | 12 | pre-treatment | 0 | SRI24 space |
| Yale | 195 | 0 | 12 | pre-treatment | 0 | SRI24 space |
| UCSF | 322 | 0 | 0 | pre- and post-treatment | 102 | Native space |
| NW | 0 | 46 | 0 | pre-treatment | 0 | SRI24 space |
| UCSD | 646 | 91 | 213 | pre- and post-treatment | 630 | Native space |
| In total | 1296 | 179 | 303 | n/a | 732 | n/a |

*Table 2.* Datasets included in the BraTS-METS 2025 Lighthouse challenge.

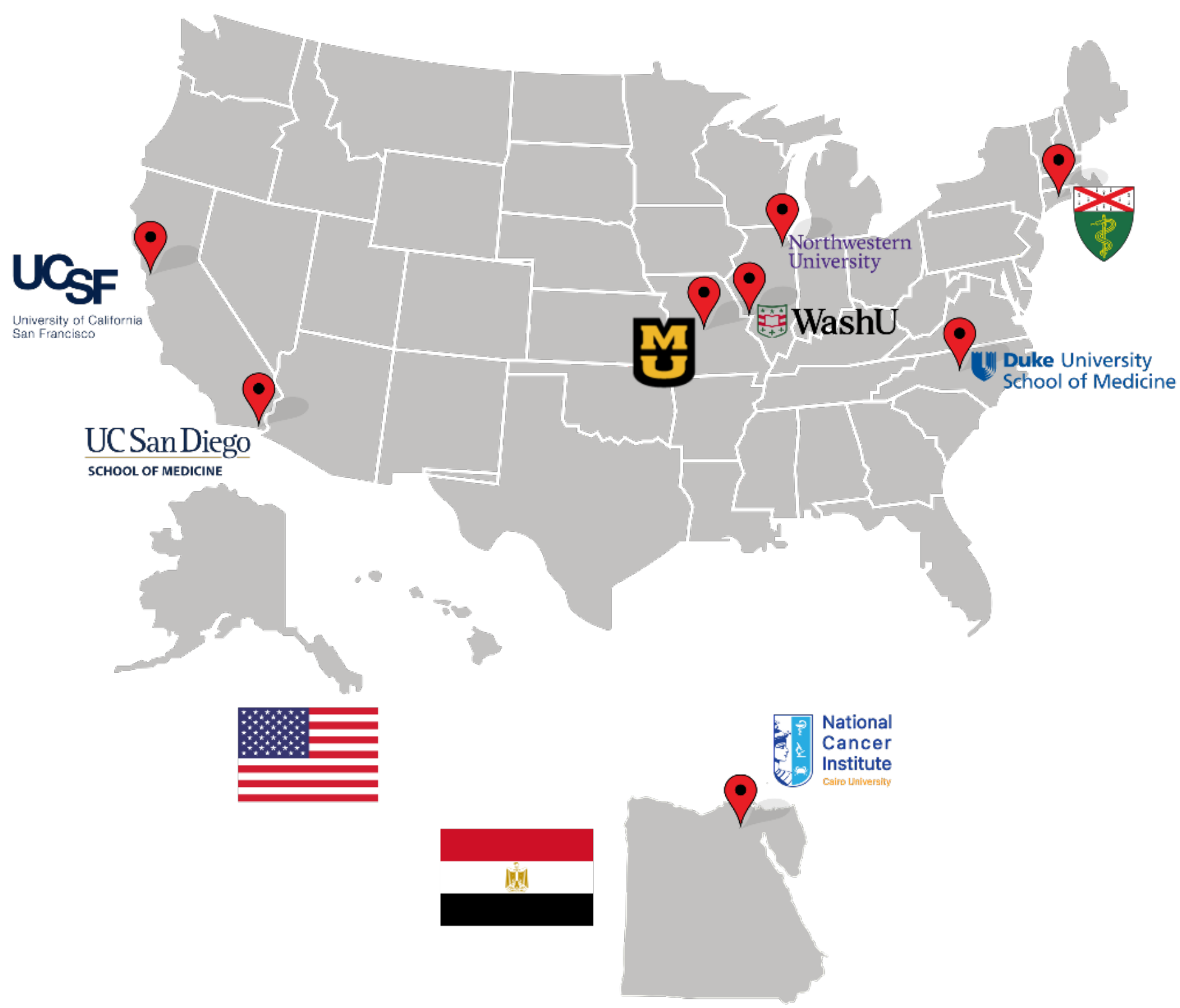

*Figure 3.* Graphical representation of the institutions that contributed the datasets for the BraTS-METS 2025 Lighthouse Challenge

**IV. Discussion:**

1. Clinical relevance of implementing AI in neuro-oncological imaging

Determination of the most appropriate treatment strategy for patients with brain metastatic disease requires detailed understanding of both the number of metastases and the gross tumor volume (GTV) and how they are changing at all timepoints of treatment regiment[30]. In addition to research indicating that GTV is inversely related to overall survival[31], clinical experience has shown that individual metastases get treated at different timepoints during patient treatment timeline, therefore having quantitative data of rate of change of each individual metastasis and how it relates to GTV is critical for personalized care of patients. These personalized treatment plans can be beneficial to treat individual metastases with better efficiency and to minimize the risk of radiation necrosis[30]. Nardone et al. found that in a cohort of patients with non-small cell lung cancer (NSCLC) brain metastases treated with SRS, a lower ratio of perilesional edema to gross tumor volume is associated with poorer overall survival[31]. Additionally, studies have shown that the volume of the surrounding edema correlates with better overall survival in patients who have undergone surgery for single brain metastasis[32]. The reported correlation of tumor and edema volumes with patient outcomes highlights the need for incorporating methods that allow tracking of different components of metastases (enhancing tumor (ET in BraTS), perilesional edema (SNFH in BraTS), and necrotic portion (NETC in BraTS)) in clinical practice that is currently only possible with implementation of AI-based segmentation methods. We hope that algorithms developed through this challenge will ensure more accurate and reproducible methods for volumetric analysis of brain metastatic disease and will lead to translation of these methods into the clinic, resulting in paradigm shift in how brain metastases are treated, which is not possible now without these technological advances.

The need for tools that assist in tracking treated brain metastatic lesions has already been emphasized in previous studies[24,33]. This process involves segmentation of brain metastases prior to treatment, re-segmentation after the treatment modality has been implied, and ultimately organizing the studies in chronological order to analyze the disease state[34]. Cassinelli-Peterson et al. evaluated a PACS-integrated tool that automatically aligns the patient's studies chronologically and provides treatment response curves and tables for individual lesions. Reporting the results in such an objective and structured way eliminates the ambiguity often found in free-text reports, which can be biased by the radiologist's personal impressions[33]. Hsu et al. developed a pipeline for automated segmentation and BM tracking across longitudinal MRI studies, demonstrating a strong correlation coefficient of 88% between the 3D longest diameter measured automatically by the model and the manual measurements taken by the radiologist[19]. Machura et al. developed a pipeline for detecting and analyzing brain metastases in longitudinal studies and concluded that it is effective and precise in tracking disease progression[35]. Kanakarajan et al. highlighted the scarcity of studies evaluating BM segmentation models on post-treatment and follow-up longitudinal data. They also underscored the fact that using 3D volume as a metric is for evaluating a lesion is superior to relying on unidimensional metrics, such as diameter, as indicators of disease progression[36].

The current gold standard for estimating treatment response in patients with brain metastases is using the Response Assessment in Neuro-Oncology for Brain Metastases (RANO-BM) criteria[37]. According to these criteria, the assessment of CNS lesion response is based on measuring longest diameter of the target lesions on axial MRI scans[38]. However, this unidimensional measurement approach can be inconsistent and suboptimal since the selected diameter might not accurately represent the actual tumor size. To better track tumor burden and changes in size over time, volumetric measurements are preferred but are currently not implemented due to lack of tools in clinical practice that would allow this. The RANO-BM group emphasizes the importance of implementing volumetric assessments in monitoring disease progression but does not require it because of the lack of availability of such tool in clinic[38]. This highlights the critical need in the field for development of tools for brain metastasis segmentation that can be translated to clinical practice. Tienda et al. studied 185 identifiable lesions in 132 patients who had undergone SRS treatment and were followed over time with MRI studies. Supplementing automated segmentation with manual refinements, the authors performed volumetric analysis on all the available pre- and post-treatment MRIs available. They concluded that that using volumetric criteria resulted in a higher sensitivity for identifying lesions as progressive, and more lesions were classified as responsive when utilizing these criteria[37]. The volumetric criteria for classifying a lesion as progressive was a minimum of 30% increase in total volume, and for a lesion to be regarded as having responded to the treatment, a minimum of 20% decrease in volume was needed[37]. All the cited studies point to a highly needed paradigm shift in how treatment response and disease progression of BM are defined. As the tools become available in clinical practice, new cutoff points can be established based on prospective clinical trials for the volumetric criteria and hopefully will be adapted by RANO-BM.

To make volumetric assessment possible, it's necessary to manually annotate a lesion for volume calculation. This task is time-consuming and tedious to perform. However, this shortcoming can be addressed through the use of algorithms that automatically segment the lesions and provide immediate volumetric data. The effectiveness of these algorithms relies heavily on the quality of ground truth data annotations. To reduce variability in manual segmentation and standardize this process, it is beneficial to involve neuroradiologists in discussions about cases where there is a lack of consensus regarding the ground truth[16]. An alternative approach would be to create a dataset annotated by more than one radiologist multiple times. It is the approach we utilized for creating the reference standard brain metastases MRI dataset in the current Lighthouse challenge. By doing this, we address the issue of inter-rater and intra-rater variability and succeed in providing a more accurate and reliable dataset.

In conclusion, continuing development of such tools for volumetric assessment of brain metastatic disease progression is crucial for improving the quality of patient follow-up assessments[39].

2. Open-science algorithms and reference standard datasets

AI-based models for brain tumor segmentation are trained and evaluated on extensive datasets[7]. Open access to high-quality annotated datasets is an essential prerequisite for creating AI tools intended to improve outcomes in neuro-oncology patients[7]. Training an AI-model on data from a single institution can limit the generalizability of the model. Therefore, data from multiple institutions, trained on a diverse population of patients, is needed to prevent data overfitting - a phenomenon that depicts the situation in which a model is over-trained on limited data and does not easily transfer to non-related institutions. In the latter case, the model has adapted to some intricacies of the small set of training data but hasn't learned the deep patterns that could render the model applicable to real clinical practice[40]. In their study, Kanakarajan et al. described two models for image segmentation on pre- and post-treatment MRI scans of patient with brain metastatic disease. The authors demonstrated that training a model on publicly available data, rather than solely on data obtained from a single institution, yields better results when the model is applied to testing data[36]. One of the aims of the BraTS-METS initiative is to develop high-quality annotated datasets that will be used to create versatile AI algorithms for imaging purposes by reducing the biases associated with the lack of open-source algorithms and open-access high-quality datasets.

3. Impact of the BraTS-METS 2025 Lighthouse Challenge

The BraTS-METS initiative is leading the way in integrating artificial intelligence into the radiological assessment of brain metastatic disease. The challenge leaders have established a framework that allows scientists from various fields - such as medicine, computer science and data science - to develop and apply deep learning algorithms for the automatic segmentation of brain metastases. This initiative manages to raise awareness about the promising potential of implementing AI algorithms in radiology to improve diagnosis, decision-making, disease progression, and treatment response assessment. The organizers of the challenge promote a culture that fosters collaboration, innovation, access to education, and fairness, which are essential pillars for advancing the role of artificial intelligence in radiology.

The challenge dataset comprising 1296 training cases with concomitant ground truth annotations and 179 validation cases is made publicly available for the participants of the challenge. After the challenge concludes, the 2025 Lighthouse multi-annotator reference standard dataset for MRI segmentation of brain metastases will be released as an open-access dataset. This effort is part of our long-term vision to establish a pre- and post-treatment brain metastasis imaging consortium for research collaboration and translation of identified algorithms into clinical practice.

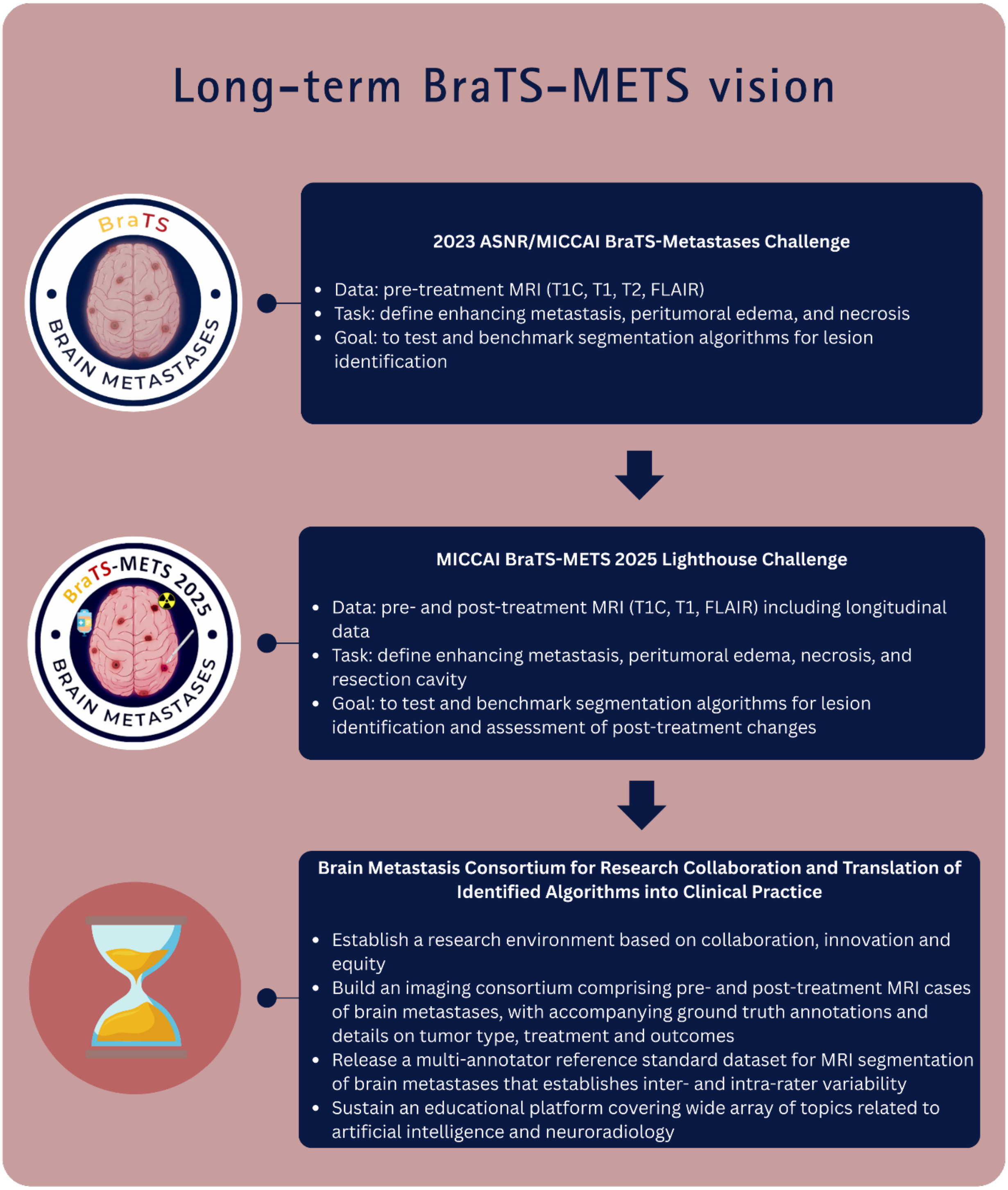

*Figure 4.* Graphical representation of BraTS-METS long-term vision for building an inter-institutional consortium for research developments based on brain MRI images of brain metastases.

4. Limitations

One limitation of the segmentation models currently used and developed for the purposes of the challenge is that they do not reliably address the issue of differentiating between tumor recurrence and pseudo-progression or radiation necrosis. After receiving radiotherapy for brain metastatic lesions, a transitory increase in enhancement volume, also known as pseudo-progression, can occur within the first 6 months following therapy. Radiation necrosis lesions within the surrounding healthy tissue can occur within 2 years after treatment and typically appear as a rim-enhancing lesion on post-contrast T1-weighed imaging, surrounded by high FLAIR signal[41]. Differentiating between the aforementioned post-treatment changes and true disease progression based solely on imaging morphology and appearance can be challenging even to experienced neuroradiologists, neurosurgeons, and neuro-oncologists[30,42]. To address this challenge, we are working on collecting clinical data information on submitted cases with information on timing after radiation treatment. This information is planned to be released as part of the consortium, with details on methods for rigorous preservation of patient information and safety currently being established.

### 5. Related studies

The scope of AI applications in the context of brain metastases is continuously expanding and underscores their increasing role in lesion detection, classification, characterization, volumetric analysis, and outcome prediction. Najafian et al. described a model that utilizes imaging features, along with radiomics, a subfield of artificial intelligence, and machine learning, to determine patterns of brain metastasis invasion[16]. Basree et al. highlight the potential of radiomics in differentiating between radionecrosis and tumor recurrence in patients who have undergone SRS for brain metastases[43]. Mouraviev et al. demonstrate promising results from incorporating radiomics in predicting radiosurgery response of BM[44]. Therefore, it is of utmost importance to conduct and facilitate initiatives like the BraTS Challenge, meant to advance the field of AI in radiology, but also to expand the scope of the BraTS-METS challenge by including functional and clinical data. We plan to expand this arm through the BraTS-METS Consortium that will address issues beyond image segmentation and will address these additional critical areas of research with a focus on collaboration and open-science philosophy.

## V. Conclusion

The BraTS-METS 2025 Lighthouse Challenge is set to play a pivotal role in advancing the capability of artificial intelligence to deliver accurate and consistent predictions for brain metastases segmentation in both pre- and post-treatment cases and for the first time establish the inter-rater and intra-rater variability in dataset reference standard annotation. The importance of conducting such a challenge is highlighted by the scarcity of literature on algorithms for BM segmentation that are used on longitudinal data comprising baseline treatment-naive MRI studies and follow-up studies obtained after the patient has undergone SRS treatment. By making use of diverse and multi-institutional datasets and ensuring that they are publicly available, along with the creation of a multi-annotated reference standard dataset, the BraTS-METS 2025 Challenge stands as a strong advocate for open-science. The immense scale of the challenge, the refined workflow pipeline, and the inclusion of longitudinal data render the BraTS-METS 2025 to be a paramount initiative providing a competitive environment for creating robust algorithms for automated segmentation of BMs and treatment response assessment in patients treated for brain metastatic disease. Our study underscores the potential of volumetric criteria for more precise response assessment and the need for more advanced AI-based algorithms that could substitute the immense time and effort required for manual 3D-segmentation of brain metastatic lesions. The goal of our team is to challenge the status quo and advance the use of AI and radiology for improving the diagnosis, management and prognosis of patients with brain metastatic disease.

**Funding**

The research reported in this article was partly supported by the Department of Radiology at Children's Hospital of Philadelphia, Philadelphia, Pennsylvania, USA.

**Conflicts of interest**

No conflicts of interest to disclose.

**Ethical considerations:**

All ethical standards, applicable laws and regulations related to treatment of human subjects are followed when conducting this study and writing this manuscript. No animals were utilized for the purposes of this research work. Institutional Review Board and Data Transfer Agreement approvals were obtained by the contributing institutions before providing the imaging data, according to regulatory standards.

**Data availability**

The datasets provided for the purposes of this challenge are available upon completion of the steps described on the following web page:
https://www.synapse.org/Synapse:syn64153130/wiki/631048
The training dataset is delivered with accompanying ground truth segmentation masks, whereas the validation dataset comes without any segmentation masks. The official testing dataset on which the challenge participants are evaluated is hidden from the participants.
The testing dataset comprising 75 brain metastasis cases that have undergone multiple stages of annotation as part of the Lighthouse initiative will be available to the public after the conclusion of the challenge.